# Revealing the behavior of soliton build-up in a mode-locked laser

Xueming Liu[*], Yudong Cui

*State Key Laboratory of Modern Optical Instrumentation, College of Optical Science and*

*Engineering, Zhejiang University, Hangzhou 310027, China*

*Corresponding author: liuxueming72@yahoo.com*

Real-time spectroscopy based on an emerging time-stretch technique can map the spectral information of optical waves into the time domain, opening several fascinating explorations of nonlinear dynamics in mode-locked lasers. However, the self-starting process of mode-locked lasers is quite sensitive to the environmental perturbation, which causes the transient behavior of laser to deviate from the true build-up process of solitons. Here, we optimize the laser system to improve its stability that suppresses the Q-switched lasing induced by the environmental perturbation. We therefore demonstrate the first observation of the entire build-up process of solitons in a mode-locked laser, revealing two possible ways to generate the temporal solitons. One way includes the dynamics of raised relaxation oscillation, quasi mode-locking stage, spectral beating behavior, and finally the stable single-soliton mode-locking. The other way contains, however, an extra transient bound-state stage before the final single-pulse mode-locking operation. Moreover, we have proposed a theoretical modeling to predict the build-up time of solitons, which agrees well with the experimental result. Our findings can bring real-time insights into ultrafast fiber laser design and optimization, as well as promote the application of fiber laser.

## I. INTRODUCTION

Transient phenomena and dynamics are the important characteristics of numerous nonlinear systems [1]-[7]. Solitons are the localized formations in nonlinear systems [8]-[15], appearing in many nonlinear processes from fluid and biology dynamics, plasma physics to fiber lasers [16]-[19], especially in mode-locked lasers [5][11][20]. The mode-locked fiber laser, as an ideal platform, is used for exploring new nonlinear phenomena due to its compact and low-cost configuration as well as excellent features of high stability and low noise. Solitons observed in mode-locked lasers exhibit several special behaviors such as soliton bunching and soliton bounding (i.e., the generation of soliton molecules) [21]-[23].

In the stationary state, the soliton train generated from mode-locked lasers can be described theoretically by means of the generalized nonlinear Schrödinger equation (NSE) [17] or Ginzburg–Landau equation (GLE) [11]. Despite the ultimate stability of the mode-locked pulse train, its initial self-starting process contains a rich variety of unstable phenomena which are highly stochastic and non-repetitive [24]-[28], and the theoretical modeling of these dynamical processes is beyond the NSE and GLE. While conventional technologies cannot generally measure these rapid non-repetitive processes due to the limited measurement bandwidth [1], the recently-developed time-stretch dispersive Fourier transform (TS-DFT) technique can provide an elegant way for real-time, single-shot measurements of ultrafast optical phenomena [29]-[34]. This technique helps scientists to experimentally resolve the evolution of femtosecond soliton molecules [30], the internal motion of dissipative optical soliton molecules [17][35], and the dynamics of soliton explosions [36]-[38]. Some successful examples of using this TS-DFT technique include the measurements of rogue wave dynamics, modulation instability, and supercontinuum generation [28][32][33].

The starting dynamics of passive mode-locking lasers has been established by a large set of experimental and theoretical investigations over the past two decades [39]-[48]. Without the emerging TS-DFT technique, the transient dynamics in mode-locked lasers had been investigated by using an oscilloscope [24][26]. The real-time spectral evolution in the build-up process cannot be resolved, however, when no TS-DFT is used [1]. Recently, both the transient spectral and temporal dynamics were observed with the assistance of the TS-DFT technique [34][35]. The environmental perturbation (e.g., the polarization change in laser cavity and the fluctuation of pumping strength) can cause the lasers to sustain the extra unstable stages, such as the Q-switched lasing [34][35]. To reveal the true build-up process of solitons, we must mitigate the environmental perturbation as far as possible. This issue is not overcome yet so far, unfortunately, although the self-starting process of mode-locked lasers had been demonstrated in the previous reports [24][26][34][35][49].

In this paper, the Q-switched lasing is availably suppressed by decreasing the environmental perturbation and optimizing the laser system with the assistance of polarization-insensitive carbon nanotube saturable absorber (CNT-SA). We therefore observe the entire build-up process of solitons in mode-locked lasers, for the first time to our best knowledge, in which there exist two possible ways to generate the laser solitons. One way comprises of the raised relaxation oscillation, quasi mode-locking (Q-ML) stage, spectral beating dynamics, and stable mode-locking. The other way includes, however, an extra transient bound-state stage prior to the final mode-locking operation. The numerical simulations based on the roundtrip circulating-pulse method confirm the experimental observations. Moreover, the theoretical modeling is proposed to investigate the raised relaxation oscillation during the build-up process of laser pulses. The build-up time for the birth of solitons is predicted, which agrees well with the experimental measurement. The entire build-up process can be theoretically described by means of the two-step method, as shown in Appendixes B and C. These findings highlight the importance of real-time measurements and provide some new perspectives into the ultrafast transient dynamics of nonlinear systems.

## II. RESULTS

### A. Build-up dynamics of solitons with beating dynamics

The time stretch technique can overcome the speed limitations of electronic digitizers [50]. The spectral interferograms are mapped into the time domain by using the TS-DFT technique. The timing data are measured via the direct detection [Fig. 1(a)], whereas, the real-time spectra are obtained by dispersing the laser pulses over a 5 km length of dispersion-compensating fiber (DCF) prior to detection [Fig. 1(b)]. The pulse-resolved acquisition and real-time spectral acquisition are measured simultaneously (see experimental setup in Appendix A). Comparing to the direct detection data of the laser output, its real-time spectral data is delayed in time by ~24.5 μs due to the additional 5 km DCF.

The direct measurement and TS-DFT data are plotted in Figs. 1(a) and 1(b), respectively. While the direct detection gives a temporal trace of the laser output as shown in panel (I), the TS-DFT technique provides the single-shot measurement of the laser spectral information as shown in panel (II). The experimental observations reveal that there exists an evidently raised relaxation oscillation before the appearance of a stable pulse train (i.e., stable mode-locking). A representative real-time measurement is exhibited in Fig. 1(b), where the duration of the raised relaxation oscillation is ~ 4.6 ms, corresponding to ~ $1.2 \times 10^5$ cavity roundtrips. Before 4.32 ms, the number of cavity photons remains at the initial low value, as determined by quantum field fluctuations [51]. From this time on, the first laser spike is generated. The separation of the neighboring laser spikes is ~ 80 μs. The build-up time of solitons from the beginning time of pumping process to the stable mode-locking is ~ 4.65 ms. Note that, in the experiments, we only need to measure the build-up period from the first laser spike to the stable mode-locking and, thus, we define the duration of this process as the nascent time of the laser (here the nascent time is ~0.35 ms).

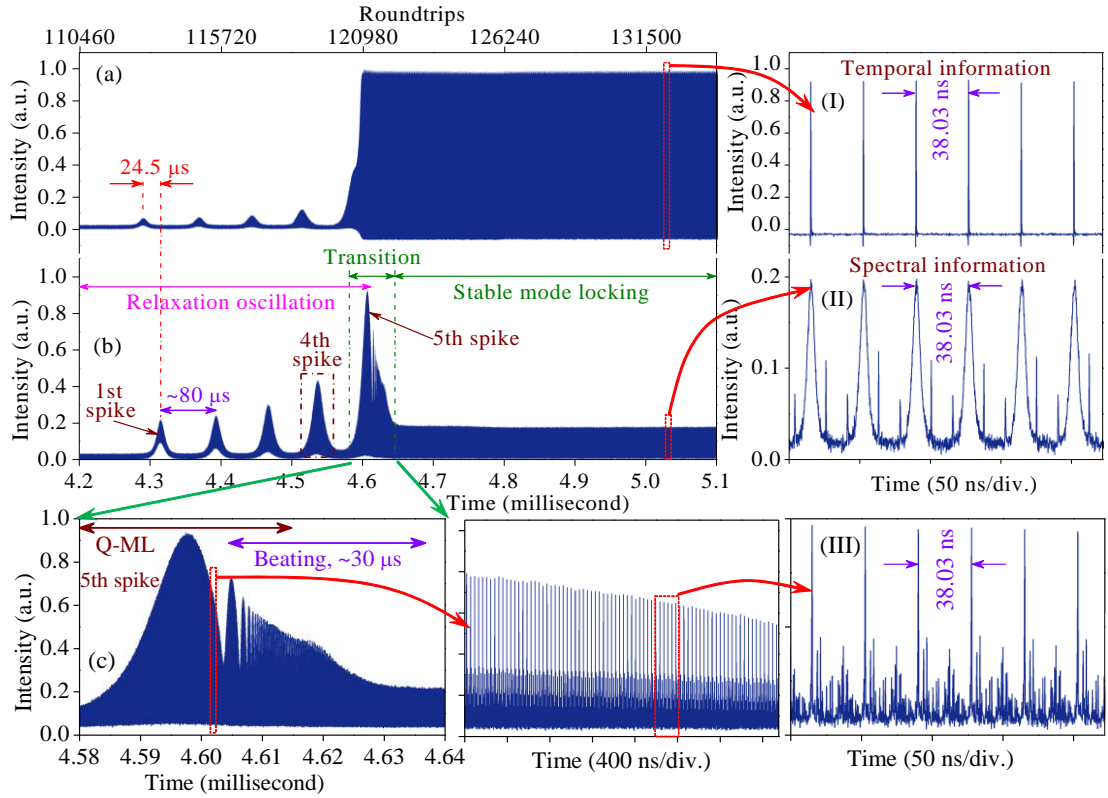

FIG. 1. Experimental real-time display of the build-up dynamics of solitons in mode-locked laser, captured simultaneously by using (a) an undispersed and (b) a dispersed element, respectively. (a) Direct detection with a high-speed photodetector and a real-time oscilloscope. (I) The expanded view shows that the timing data represent the temporal information. (b) The build-up dynamics is obtained by dispersing the solitons in 5 km DCF prior to detection, as denotes the single-shot spectral information shown in (II). The timing data advance the real-time spectral data about 24.5 μs that is delayed by 5 km DCF. (c) Close-up of the data from (b) for the 5th laser spike, revealing the quasi mode-locking (Q-ML) and beating dynamics with the duration of ~ 30 μs. (III) There are multiple subordinate pulses together with a dominant pulse during the laser cavity. The period for (I) to (III) is 38.03 ns, corresponding to the roundtrip time of laser cavity. The build-up process of mode-locked lasers includes the raised relaxation oscillation, Q-ML stage, beating dynamics, and stable mode-locking.

Usually, the build-up process of mode-locked lasers includes the raised relaxation oscillation, Q-ML stage, beating dynamics, and finally the stable mode-locking state, as shown in Fig. 1. The experimental results show that the laser spikes are generated gradually. Figure 1(c) shows the expanded views of the 5th laser spike in Fig. 1(b). Figure 1(c) exhibits that, prior to the stable mode-locking, there exists a Q-ML and beating dynamics with the duration of ~ 30 μs. During the Q-ML stage, multiple subordinate pulses appear together with a dominant pulse as shown in panel (III). Panel (II) reveals that the TS-DFT technique can achieve spectral information rather than temporal information on ultrashort time scale. But, the direct detection without the TS-DFT technique only is used to obtain the temporal information, as shown in panel (I). The panels (I) to (III) show that the roundtrip time of laser is 38.03 ns.

The time-continuous data stream shown in Fig. 1(b) evolves with a periodicity of 38.03 ns that corresponds to the cavity roundtrip time. We therefore segment the time series into intervals of 38.03 ns, achieving a two-dimensional (matrix) representation exhibited in Fig. 2(a). Here, the vertical and horizontal axes depict the information within a single roundtrip and the dynamics across consecutive roundtrips, respectively. The two-dimensional representation (also named as spatio-temporal representation) had been proposed to describe the nonlinear system with delayed feedback [52], and the analogous techniques had been used in several optical systems, such as mode-locked fiber laser [53], cavity soliton in fiber-ring resonator [54], and injection-locked semiconductor laser [55].

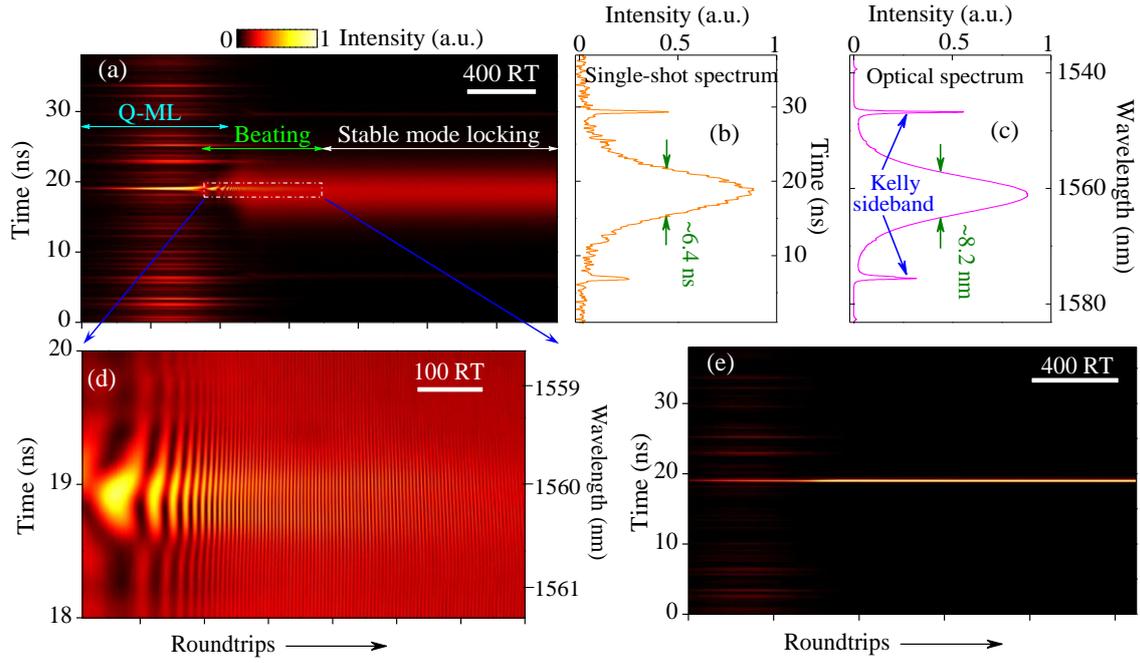

FIG. 2. Formation of a soliton with beating dynamics. The recorded time series is segmented with respect to the roundtrip time and displays the build-up dynamics of a soliton. The intensity profile evolves along with time (vertical axis) and roundtrips (horizontal axis). (a) Experimental real-time observation during the formation of a soliton from the Q-ML and beating behavior to the stable mode locking (see Supplemental Material [56] for the full animation). The experimental data are from Fig. 1(b). TS-DFT maps the spectral information into the temporal domain. (b) Exemplary single-shot spectrum, corresponding to the last frame in (a). (c) Optical spectra of soliton measured by an optical spectrum analyzer (OSA). (d) Close-up of the data from (a), revealing the interference pattern for the beating dynamics. (e) Experimental real-time observation via the direct measurement (see Supplemental Material [60] for the full animation), no using TS-DFT technique. The beating dynamics is not revealed in the direct measurement. RT, roundtrip (scale bar).

Figure 2(a) exhibits a complex formation process of the pulse laser, involving Q-ML stage, beating dynamics, and ultimately the stationary single-soliton mode-locking state. A video illustrates the experimental real-time observation in detail in the Supplemental Material [56]. There exist multiple pulses in the laser cavity during the Q-ML stage and the beating dynamics, whereas only a dominant pulse gradually evolves to the final stationary mode-locking pulse. The last frame in Fig. 2(a) is illustrated in Fig. 2(b). The corresponding optical spectrum is demonstrated in Fig. 2(c) that is measured directly by an optical spectrum analyzer (OSA). Clear Kelly sidebands are observed in Fig. 2(c), being the typical characteristics of soliton fiber lasers [57]-[59]. The full-width-at-half-maximum (FWHM) spectral width of the solitons is ~8.2 nm. Figure 2(d) illustrates the close-up of the data from Fig. 2(a), revealing the beating dynamics with an interference pattern. Experimental real-time measurements from undispersed events (i.e., no using TS-DFT) are demonstrated in Fig. 2(e), where the beating phenomenon cannot be discovered (see the animation in the Supplemental Material [60]).

**B. Build-up dynamics of solitons with transient bound state**

Figures 1 and 2 show that the beating dynamics occurs prior to the stable mode-locking. Some of our experimental observations demonstrate, however, that a transient bound state of two solitons may appear between the two stages. A typical example is illustrated in Fig. 3 and the corresponding animation is included in Supplemental Material [61] for details. The real-time TS-DFT measurement of this unique build-up process is shown entirely in Fig. 3(a), which includes the raised relaxation oscillation, the transition region, and the final stationary single-pulse mode-locking state. The unique transition region shown in Fig. 3(a) has a duration ~3 times longer than that in Fig. 1(b). Figure 3(b), which is quite similar to panel (II) in Fig. 1, is the expanded view at the stable mode-locking in Fig. 3(a). Figure 3(c) is an enlarged plot that demonstrates the evolution of optical wave during this transition region in detail. The interferograms show the periodical modulation along wavelength, which is the typical result of bound-state spectrum [17][30]. Figure 3(e) demonstrates the Fourier transform of each single-shot spectrum of the transient bound state. Obviously, the corresponding field autocorrelation with three peaks exhibits the evolution of bound state with two pulses [30].

We set the beginning time of the stationary single-pulse mode-locking state as 0 roundtrip number for the convenience of reference, as shown in Fig. 3(c). For example, the roundtrip number of -800 corresponds to the recording time 800 roundtrips prior to the stable mode-locking state. Figure 3(c) shows that the two solitons of transient bound state are generated via the beating effect at the roundtrip of about -2800, where the separation of solitons is zero [Fig. 3(e)]. During the transition, the two solitons evolve in a rather complex manner, and finally the bound state collapses due to the increasing separation between them. The single-shot spectrum at the steady state is similar to Fig. 2(b). The relative phase, $\alpha$, between the two solitons experiences eight abrupt changes during the roundtrips from -1840 to -710 [see Fig. 3(f)], which is revealed by the eight circles in the interaction plane shown in Fig. 3(d). The relative phase evolves on the opposite direction after the roundtrip number of about -1100 [Fig. 3(f)] so that the rotating direction reverses after this inflection point [Fig. 3(d)].

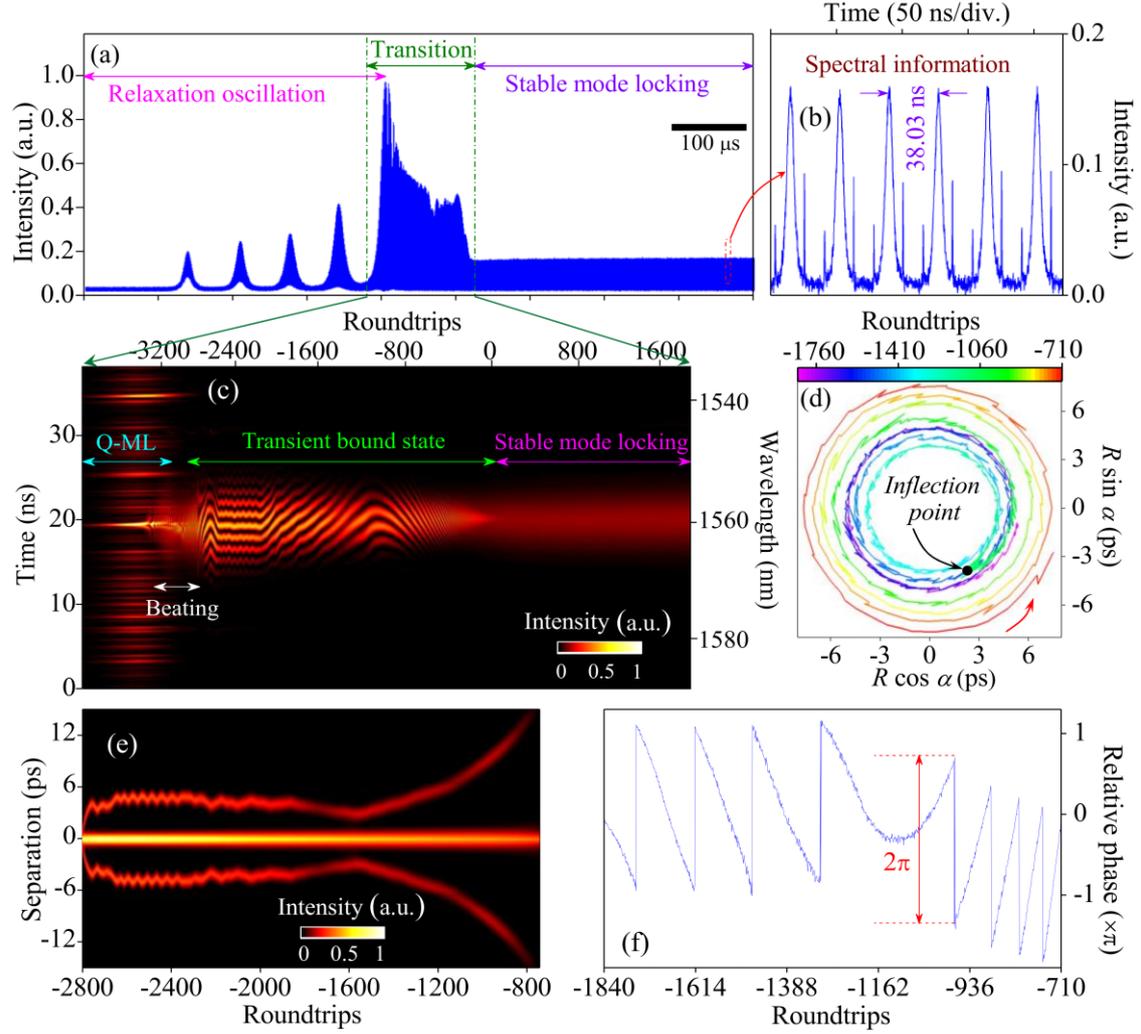

FIG. 3. Experimental real-time display of the build-up dynamics of solitons with transient bound state. (a) Build-up dynamics via the TS-DFT technique. The entire build-up process includes the raised relaxation oscillation, transition region, and stable mode-locking. (b) Close-up of the data from (a) at the stable mode-locking state, as denotes the single-shot spectral information. (c) Experimental real-time interferograms during the formation of a soliton, accessed via the dynamics of Q-ML, beating dynamics, transient bound state, and stable mode-locking. For a full animation of the observations, see Supplemental Material [61]. For the convenience of reference, the beginning time of stable mode-locking is set to be zero roundtrip number. (d) Dynamics of the soliton mapped in the interaction plane over 1000 roundtrips. The angle, $\alpha$, represents the relative phase of two solitons. The radius, $R$, corresponds to the bound-state separation. Two solitons gradually depart from each other. (e) The Fourier transform of each single-shot spectrum corresponds to a field autocorrelation of the momentary bound state, tracing the separation between both solitons. (f) The relative phase between both solitons along with roundtrips.

Figure 3(c) exhibits the spectral interference since the TS-DFT technique is used to obtain spectral information rather than temporal information. The Fourier transform of each single-shot spectrum only achieves the field autocorrelation of the momentary bound state [30], as shown in Fig. 3(e). To extract the timing information of transient bound states, we assume a bound state as a superposition of temporally separated individual solitons. The bound state field can be expressed as [17][30]

$$E(t) = \left[ E_1(t) + E_2(t)\exp(i\alpha) \right] \exp(i\omega_0 t). \quad (1)$$

Here $\omega_0$ is the common carrier frequency. The temporal solitons usually have the hyperbolic secant envelopes [42][59][62], i.e.,

$$E_1(t) = a_1 \text{sech}\left[ (t - R/2)/\Delta t_1 \right], \quad (2)$$

$$E_2(t) = a_2 \text{sech}\left[ (t + R/2)/\Delta t_2 \right], \quad (3)$$

where $a_1$ and $a_2$ are the amplitudes of the two solitons in the transient bound state, respectively, and $\Delta t_1$ and $\Delta t_2$ are the corresponding pulse widths. After the manipulation of Fourier transform, the spectral intensity $I(\Omega)$ of the optical field is given by

$$I(\Omega) = A_1^2 \text{sech}^2\left( \frac{\pi}{2}\Delta t_1 \Omega \right) + A_2^2 \text{sech}^2\left( \frac{\pi}{2}\Delta t_2 \Omega \right) + 2A_1 A_2 \text{sech}\left( \frac{\pi}{2}\Delta t_1 \Omega \right) \text{sech}\left( \frac{\pi}{2}\Delta t_2 \Omega \right) \cos(\Omega R + \alpha), \quad (4)$$

where $A_1 = \sqrt{\pi/2} a_1 \Delta t_1$, $A_2 = \sqrt{\pi/2} a_2 \Delta t_2$, and $\Omega = \omega - \omega_0$.

On the basis of the experimental data shown in Fig. 3, $a_1$, $a_2$, $\Delta t_1$, and $\Delta t_2$ can be achieved by solving Eq. (4) with a nonlinear least-square solver. Then the temporal intensity, $|E(t)|^2$, for the bound state field is demonstrated in Fig. 4. It can be seen that the two solitons in the transient bound state have different amplitudes and pulse widths. The two solitons are generated quickly from the roundtrip of around -2800 and, then, their separation fluctuates within a range of ~5.5 ps before the roundtrip number of about -1250. In succession, the two solitons start to depart from each other.

Finally, one of the solitons disappeared, leaving the other one evolving to the stationary mode-locking pulse inside the laser cavity. The system ultimately settles on the stable mode locking with only one soliton.

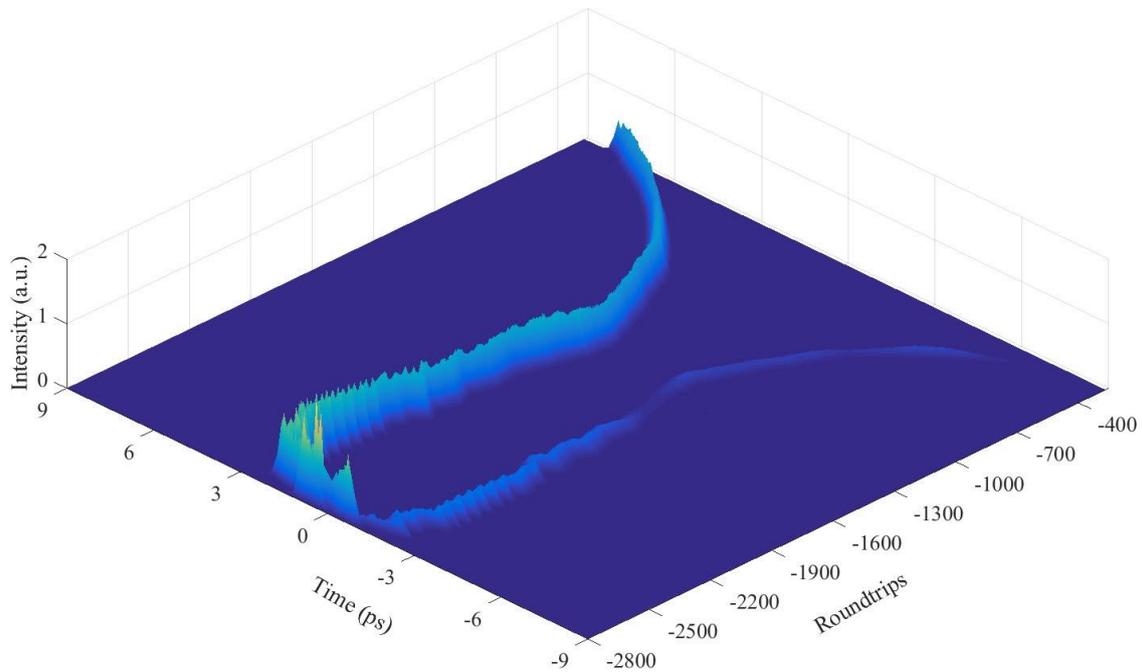

FIG. 4. Interaction and evolution of two solitons in the transient bound state. The temporal solitons are extracted from the experimental data shown in Fig. 3 by using the levenberg-marquardt algorithm. The stronger soliton gradually evolves to the stationary state, however, the weaker one ultimately vanishes via the complex dynamics. The roundtrips are from about -2800 to -300, and the corresponding autocorrelation intensity of the momentary bound state is shown in Fig. 3(e).

## III. DISCUSSIONS

**A. Build-up process of mode-locked lasers with Q-switched lasing**

Kerr-lens mode-locked laser is able to deliver the pulses and solitons [1][30], in which the nonlinear effect plays a critical role. To simplify the system and improve the stability of lasers, the mode-locked fiber lasers had been proposed on the basis of the nonlinear polarization rotation (NPR) technique [34][49]. However, the experimental observations show that the self-starting process of mode-locked laser is quite sensitive to the environmental perturbation, such as the instability of laser diode, the polarization change in laser cavity, and the fluctuation of pumping strength. Therefore, the change of polarization state in laser can extend the build-up time of mode-locked lasers. For instance, the build-up time of lasers is more than 10 ms [1] and, even, more than 100 ms [35].

We establish a fiber laser mode-locked by NPR technique, in which the polarization change and the fluctuation of pumping strength can evidently influence the self-starting process. Figure 5(c) shows a typical build-up process of mode-locked laser. Obviously, the Q-switched lasing occurs prior to the stable mode-locking. The number of Q-switched lasing is 189 with the duration of more than 160 ms. This result is very similar to the experimental observation reported in [35].

We improve the stability of pumping strength and decrease the environmental perturbation. The build-up time of laser is shortened to be about 80 ms with 76 lines of Q-switched lasing, as shown in Fig. 5(b). We further enhance the robustness of mode-locked laser by using the hybrid saturable absorber where the single-wall carbon nanotubes together with NPR technique operate on the laser. The experimental result is demonstrated in Fig. 5(a), where the duration of Q-switched lasing is ~3 ms and its number is 5. Every Q-switched lasing is composed of several hundreds of pulses, as shown in the left inset of Fig. 5(a).

It can be seen that, by comparing Fig. 5(a) to Figs. 5(b) and 5(c), the decrease of environmental perturbation can effectively shorten the build-up time of laser and suppress the Q-switched lasing. Fig. 5(a) shows that, therefore, the laser experiences a shorter unstable Q-switched lasing stage. Note that, within Fig. 5(a), the separation between the neighboring Q-switched lasing lines is ~700 μs. However, the separation between the laser spikes in the raised relaxation oscillation is ~80 μs, as shown in Fig. 1(b).

When the polarization-dependent devices are removed from the laser cavity, only single-wall carbon nanotube serves as the saturable absorber. Simultaneously, we optimize the laser system and the saturable absorber. As a result, the Q-switched lasing is suppressed completely and, then, no Q-switched lasing occurs in the self-starting process of mode-locked laser. The raised relaxation oscillation is observed experimentally, as shown in Fig. 6(a) below. Therefore, the experimental results here denote the general build-up process, which can reflect the intrinsic features of mode-locked lasers.

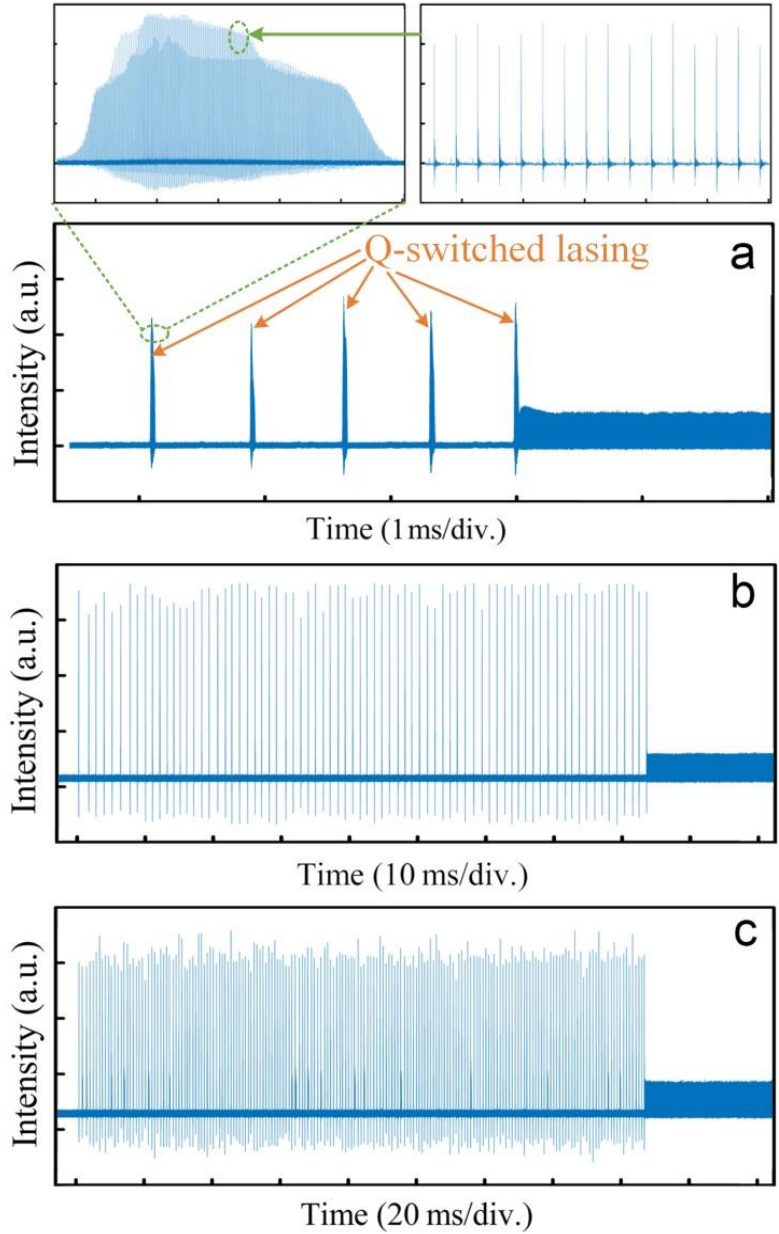

FIG. 5. Build-up process of mode-locked laser with Q-switched lasing. The laser experiences an unstable Q-switched lasing stage during the build-up process. (a) Hybrid saturable absorber based on the single-wall carbon nanotubes together with NPR technique. The duration of Q-switched lasing is about 3 ms with 5 lines. The separation between the neighboring Q-switched lasing lines is ~700 μs. Inset (Left): close-up of the data from (a) for a Q-switched lasing. Inset (Right): expanded view of a part from the left inset figure. (b) NPR-based saturable absorber together with weaker fluctuation of pumping strength. The duration of Q-switched lasing is more than 80 ms with 76 lines. (c) NPR-based saturable absorber together with stronger fluctuation of pumping strength. The duration of Q-switched lasing is more than 160 ms with 189 lines.

**B. Build-up process of mode-locked laser without Q-switched lasing**

Figure 5 shows that the Q-switched lasing can cover up the true build-up dynamics of mode-locked lasers. The polarization-independent modelocker (e.g., single-wall carbon nanotubes) can availably suppress the Q-switched lasing after the laser system is optimized. A typical result for the build-up process is demonstrated in Fig. 6. Figures 6(a) and 6(b) show the experimental results and the theoretical predictions, respectively. The red and blue curves denote the evolution of the pumping rate and raised relaxation oscillation along with time, respectively. The modeling for describing the raised relaxation oscillation is shown in Appendix B.

The relaxation oscillation is a general transient behavior of laser [51]. The laser without a modelocker undergoes a damped oscillatory behavior and, finally, approaches to the stable continuous-wave (CW) operation [63]. In contrast, the modelocker-based laser experiences a raised relaxation oscillation rather than a damped behavior, as shown in Fig. 6. Such relaxation oscillation was not observed in the previous reports [24][26][34][35][49], since the Q-switched lasing leads to the laser departing from such phenomenon. The modelocker can drive many longitudinal modes to be from randomized to synchronized, where the phase difference $\theta$ between any two neighboring modes is gradually locked to a constant value. During the raised relaxation oscillation, $\theta$ is a variable rather than a constant. We therefore define the former (i.e., a variable for $\theta$) as Q-ML and the latter (i.e., a constant for $\theta$) as perfect or stable mode-locking. The transition stage from Q-ML to stable mode-locking undergoes a beating dynamics [see Fig. 2(a)], which is described as auxiliary-pulse mode locking by Herink *et al.* [1]. Moreover, both beating and transient bound state stages occur at the transition from Q-ML to stable mode-locking (see Fig. 3), which is observed here for the first time.

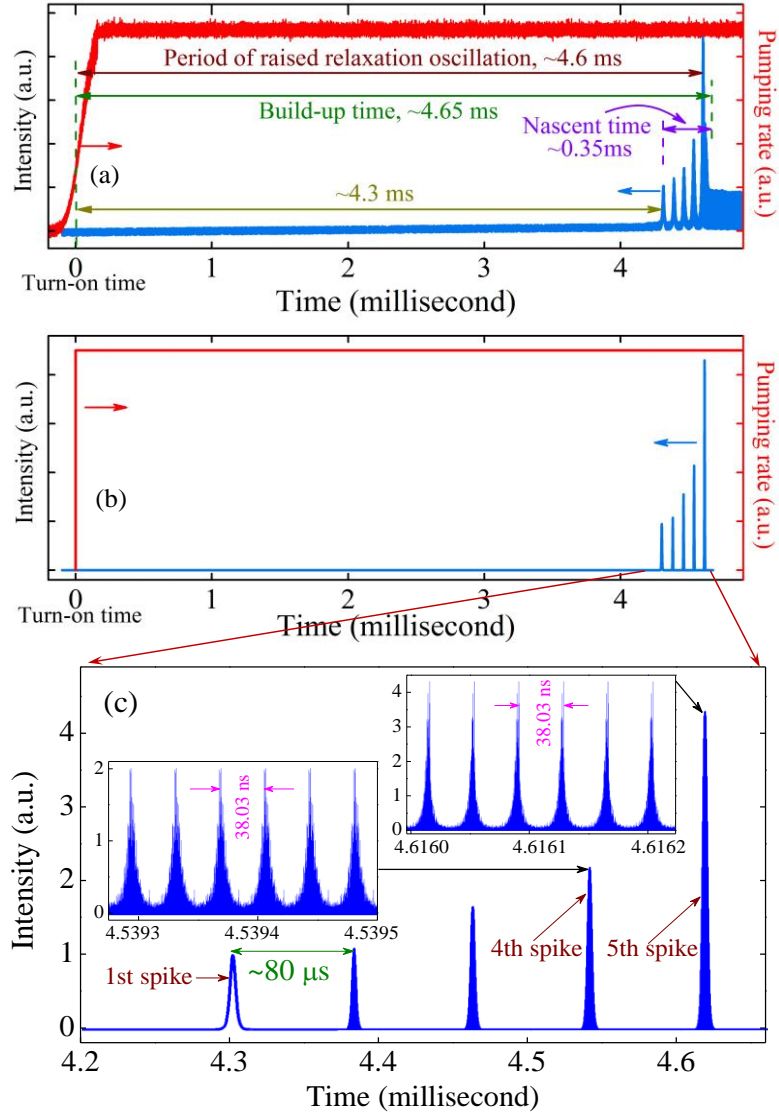

FIG. 6. Build-up process of mode-locked laser without Q-switched lasing. (a) Experimental results. (b) Numerical simulation, corresponding to (a). To numerically simulate the raised relaxation oscillation, we assume that the phase difference $\theta$ between any two neighboring modes is uniformly distributed from -4.8 to 4.8, from -4.2 to 4.2, from -3.65 to 3.65, and from -2.86 to 2.86 for the second to fifth laser spikes, respectively. Red curves indicate the evolution of pumping rate along with time. Blue curves show the evolution of raised relaxation oscillation. (c) Close-up of (b). Two insets in (c) are the expanded view of the 4th and 5th laser spikes. Note that a.u. denotes arbitrary unit. The raised relaxation oscillation at the beginning stage of the mode-locking can be precisely predicted with several key features, matching excellently with the experimental measurements.

The build-up time is defined as the duration from the beginning time (i.e., turn-on time) of pumping process to the stable mode-locking. The nascent time is defined as the duration from the first laser spike to the stable mode-locking. The period of raised relaxation oscillation is defined as the duration from the turn-on time to last laser spike. Therefore, the build-up time and nascent time are ~4.65 and 0.35 ms, respectively, as shown in Fig. 6(a). The first laser spike occurs at ~4.3 ms and the period of raised relaxation oscillation is ~4.6 ms. Both the theoretical prediction and experimental observation show that, although the pump power is inputted into the laser system after the turn-on time, the number of cavity photons remains at the initial low value before the first laser spike (i.e., ~4.3 ms). Therefore, no pulse appears before ~4.3 ms, as shown in Fig. 6.

## C. Theoretical confirmation

The proposed modeling in Appendix B successfully simulates the relaxation oscillation process, but it cannot describe the dynamics and evolution of pulses in the stationary mode-locking. Based on the extended NSE and the roundtrip circulating-pulse method (see Appendix C), we can numerically achieve the two different evolution processes of soliton build-up. Figure 7 demonstrates two typical results of numerical simulations, revealing two different evolution ways of the soliton build-up. In one way, the single-soliton operation is directly formed via the beating dynamics, as shown in Figs. 7(a) and 7(b). Figures 7(a) and 7(c) exhibit the spectral and temporal evolutions of pulses, respectively. The simulation starts at an initial signal with the noise background that is illustrated as the red curve in Fig. 7(g). Obviously, the numerical results are in good agreement with the experimental observations shown in Fig. 2. When the simulation starts with another noise background field [see the black curve in Fig. 7(g)], the transient bound state can be observed in the build-up process, as shown in Fig. 7(d) (i.e., spectral evolution) and Fig. 7(f) (i.e., temporal evolution). In this way, the transient bound state with two solitons is generated at the beginning and, however, the stronger soliton gradually evolves to the stationary state and the weaker one slowly decays and ultimately vanishes via the complex dynamics. The temporal evolution of two pulses in the numerical simulation [i.e., Fig. 7(f)] is quite similar to the experimental results (i.e., Fig. 4). Figures 7(b) and 7(e) illustrate the close-ups of the white dotted-line boxes in Figs. 7(a) and 7(d), respectively. However, the finally stable solutions for two build-up ways have the same values, as shown in Fig. 7(h) (i.e., spectral domain) and Fig. 7(i) (i.e., temporal domain). It should be noted that the two build-up ways are achieved at the same laser parameters except that two initial noises [see Fig. 7(g)] have the slight difference.

The roundtrip circulating-pulse method together with NSE (see Appendix C) can be used to simulate the two different evolution processes of soliton build-up and achieve the Kelly sidebands numerically, as shown in Fig. 7. However, they cannot simulate the raised relaxation oscillation at the beginning of the birth of solitons. A detailed description and analysis are given in the Supplemental Material [64]. On the other hand, the proposed modeling in Appendix B can simulate the raised relaxation oscillation process and predict the build-up time and the nascent time to be ~4.65 and 0.35 ms [see Fig. 6(b)], respectively. But it cannot achieve the dynamics and evolution of solitons (e.g., Fig. 7). Here, by combining Appendix B and C, we have proposed a two-step method to simulate the entire build-up process of solitons in mode-locked lasers, including the raised relaxation oscillation, transient bound state, Kelly sidebands, and beating pattern. The proposed method is described in the Supplemental Material in detail [64].

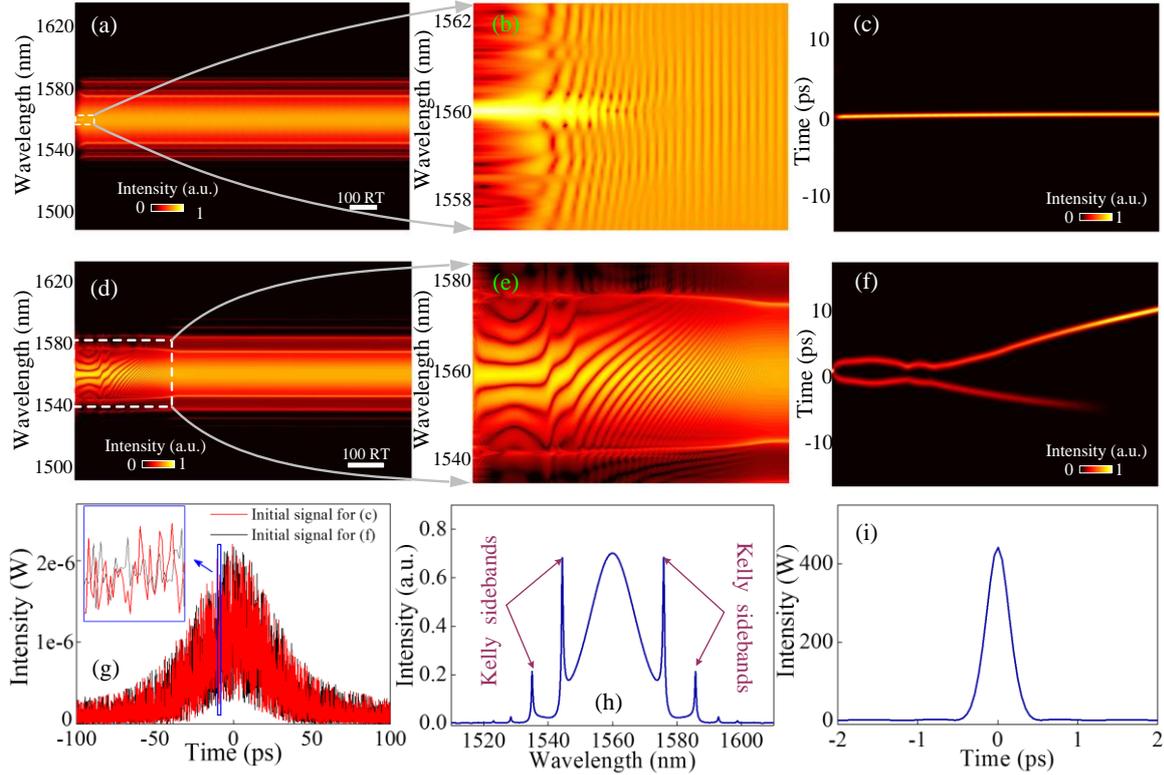

FIG. 7. Numerical simulations based on the roundtrip circulating-pulse method. (a)-(c) Soliton build-up process with beating dynamics in one way. (d)-(f) Soliton build-up process with transient bound state in another way. (a) and (d) Spectral evolution of soliton along with roundtrips. (c) and (f) Temporal evolution of soliton along with roundtrips. (b) and (e) Close-ups of the data from the white dotted-line boxes in (a) and (c), respectively. (g) Initial signals with the noise background. The intensity of initial signals is in the order of magnitude of $10^{-6}$ W, which is eight orders of magnitude lower than the intensity of stable solitons (i.e., $10^2$ W). Inset: close-up of the data from the blue box. (h) Spectral and (i) temporal profiles at the finally stationary solution. Although the solitons experience through two different build-up ways, they evolve to the same steady soliton solution with Kelly sidebands.

## V. CONCLUSIONS

The Q-switched lasing covers up some important phenomena in the build-up process of mode-locked lasers and leads to the self-starting time to be quite long (see Fig. 5). By optimizing the laser system and the modelocker, we suppress completely the Q-switched lasing that is induced by the environmental perturbation. The experimental results show that, without the Q-switched lasing, the build-up time of mode-locked lasers is ~4.65 ms and the nascent time of solitons is ~0.35 ms [see Figs. 1(b) and 6(a)]. We experimentally observe the real-time spectral dynamics of the entire build-up process of solitons, for the first time to our best knowledge, by using the TS-DFT technique. It is discovered that there are two ways for the build-up process of solitons in mode-locked lasers. The entire build-up process usually includes the raised relaxation oscillation, Q-ML stage, and beating dynamics in one way (see Fig. 1), and even contains an extra dynamics such as the transient bound state of two solitons in another way (see Fig. 3). We have proposed the modeling for describing the raised relaxation oscillation in the birth of solitons (see Appendix B), which can successfully predict the build-up time of solitons (see Fig.6). Based on this modeling together with the extended NSE and roundtrip circulating-pulse method (see Appendix C), we have proposed the two-step method that can accurately simulate the entire build-up process of soliton. These findings provide new perspectives into the ultrafast transient dynamics and bring real-time insights into laser design and applications [65]. Such real-time spectroscopy technique is expected to provide new insight into a wider class of phenomena in complex nonlinear systems [30]. For more application prospects of our work, please see Supplemental Material [64].

**ACKNOWLEDGMENTS**

We thank X. Yao, X. Han, G. Chen, W. Li, G. Wang, and Y. Zhang for fruitful discussions. The work was supported by the National Natural Science Foundation of China under Grants No. 61525505, No. 11774310, and No. 61705193, by the Key Scientific and Technological Innovation Team Project in Shaanxi Province (2015KCT-06), and by China Postdoctoral Science Foundation (2017M610367).

**APPENDIX A: EXPERIMENTAL SETUP**

The schematic diagram of mode-locked laser is shown in Fig. 8. The laser system consists of a CNT-SA, a 3.5-m-long erbium-doped fiber (EDF) with 6 dB/m absorption at 980 nm, a polarization controller (PC), some single-mode fiber (SMF) pigtails, and a polarization-independent hybrid combiner of a wavelength division multiplexer, tap coupler and an isolator (WTI). The polarization-independent WTI is used to ensure the unidirectional operation, extract intracavity power with a ratio of 10%, and input the pump power from the laser diode (LD). An optical chopper is placed in the free-space section between LD and WTI to control the onset of solitons. CNT film acts as a modelocker to initiate the soliton operation. PC is utilized to optimize the mode-locking performance by adjusting cavity linear birefringence. The length of SMF is ~ 4.3 m. EDF and SMF have dispersion parameters of about 11.6 and −22 $ps^2$/km at 1550 nm, respectively. The total dispersion of laser cavity is about -0.06 $ps^2$ and the total cavity length is ~ 7.8 m.

The real-time temporal and time-averaged spectral detections for solitons are recorded with a high-speed real-time oscilloscope and an OSA, respectively. The spectral information is mapped into the temporal waveform via DCF. The length of DCF is ~5 km with the dispersion of about -160

ps/(nm km).

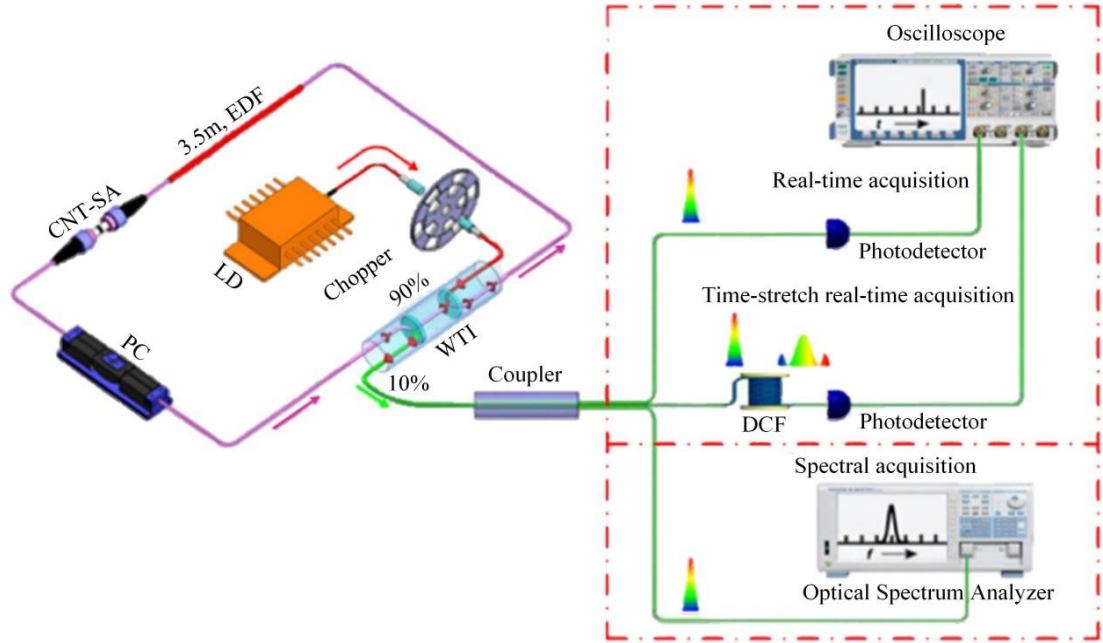

FIG. 8. Schematic diagram of the experimental setup for the mode-locked laser, generating the temporal solitons. The output can be characterized with real-time acquisition or time-averaged spectral acquisition (i.e., via optical spectrum analyzer). Real-time acquisition harnesses a high-speed real-time oscilloscope (20 GSa/s sampling rate) together with a high-speed photodetector. Timing data are achieved with direct detection and real-time spectral data are obtained by dispersing the solitons in dispersion-compensating fiber (DCF) prior to detection. EDF, erbium-doped fiber; PC, polarization controller; LD, laser diode; CNT-SA, carbon nanotube saturable absorber; WTI, hybrid combiner of wavelength division multiplexer, tap coupler and isolator.

## APPENDIX B: RELAXATION OSCILLATIONS FOR MODE-LOCKED LASER

The Kerr-lens lasers cannot usually self-start [1]. In contrast, the laser here is mode-locked by the single-wall carbon nanotubes that serve as an excellent saturable absorber (i.e., modelocker) to self-start operation. The experimental setup is described in above in detail. Some transient features at the nascent stage of this laser can be expressed by the rate equations, which provide a rather simple and intuitive picture of the lasing behavior. They can be written as [51][63]

$$\frac{du}{dt} = R_p - uB_s q - \frac{u}{T_u}, \tag{B1a}$$

$$\frac{dq}{dt} = uB_s q - \frac{q}{\xi} + \frac{u}{p_m T_u}. \tag{B1b}$$

Here $u$ and $q$ represent the population inversion and the photon number in the laser cavity, respectively. $R_p$ is the pumping rate. $T_u$ is the lifetime of the upper-laser level. $\xi$ is the photon lifetime related to the cavity loss. $B_s$ is the Einstein coefficient given by $B_s = 1/(p_m T_u)$. $p_m$ is the number of cavity modes coupled to the fluorescent line and is given by $p_m = 8\pi v^2 \Delta v V / c^3$, where $v$ is the carrier frequency of photons, $\Delta v$ is the bandwidth of the laser medium, $c$ is the speed of light in vacuum, and $V$ is referred to as the mode volume within the laser cavity.

The total cavity loss, $\varsigma$, is time independent when the modelocker is absent in the laser cavity. Then $\xi$ is a constant given by $\xi = L/(c\varsigma)$, where $L$ is the optical length of the laser resonator. Equation (B1) can be used to describe the laser behaviors of both the CW and the transient operation, giving sufficiently accurate results for most practical purposes. On the contrary, $\varsigma$ is time-dependent when the modelocker (e.g., carbon nanotubes) is present in the laser cavity. The corresponding equation is expressed by [66]

$$\varsigma = \varsigma_0 + \alpha_0 / (1 + I / I_{sat}), \tag{B2}$$

where $\alpha_0$, $\varsigma_0$, and $I_{sat}$ are the linear limit of saturable absorption, nonsaturable loss, and saturation intensity, respectively. $I$ represents the input optical intensity which is related to the photon number $q$ in the laser cavity. The relationship is given as $I=qchv/L$, where $h$ is Planck constant [51].

For deriving Eq. (B1), we assume that the laser is oscillating on only one cavity mode [51]. In the mode-locking regime, actually, the laser with the cavity length of ~ 7.8 m generates pulses with the spectral bandwidth of ~8.2 nm, corresponding to about 40000 locked modes. The stable mode-locking stage occurs when phases of various longitudinal modes are synchronized such that the phase difference between any two neighboring modes is locked to a constant value [62]. Usually, pulses during the build-up process would not satisfy the ideal condition [67]. Therefore, the deviations of the phase from the ideal values had been proposed by introducing a fluctuating background [68]. During the self-starting process of the pulse laser, the modelocker can gradually drive thousands of modes from absolutely random state to the stable mode-locking state that these cavity modes have regular phase relationships. The total optical field in the laser cavity can be expressed as [69]

$$A(t) = \sum_{m=-M}^{M} A_m \exp(i\theta_m - i\rho_m t) , \qquad (B3)$$

where $A_m$, $\theta_m$, and $\rho_m$ are the amplitude, phase, and frequency of a specific mode among 2M+1 modes, respectively. The phase difference between any two neighboring modes is defined as $\theta = \theta_m - \theta_{m-1}$.

When the modelocker is present in the laser cavity, the laser parameters are employed in our simulations, including $R_p$=3.75×10$^{16}$ s$^{-1}$ (corresponding to the pump power of ~ 16 mW), $T_u$=10×10$^{-3}$ s, $\Delta\nu$=4.68×10$^{12}$ s$^{-1}$, $V$=9×10$^{-10}$ m$^3$, $\alpha_0$=4.302%, and $I_{sat}$=9.67 MW/cm$^2$. $\theta$ is uniformly distributed on the interval from -$\theta_0$ to $\theta_0$. $\theta_0$ is assumed as 4.8, 4.2, 3.65, and 2.86 for the second to fifth laser spikes, as shown in Fig. 6(c), respectively. The phase differences of modes for the first laser spike distribute absolutely randomly. The simulation results are shown in Fig. 6(b), where Fig. 6(c) is the enlargement of all laser spikes. Each spike contains ~ 300 cavity roundtrips. Two insets in Fig. 6(c) are the local enlargements of the 4th and 5th laser spikes.

APPENDIX C: ROUNDTRIP CIRCULATING-PULSE METHOD FOR MODE-LOCKED LASER

Although Eqs. (B1)-(B3) can successfully predict the raised relaxation oscillation, they cannot describe the evolution and interaction of solitons in the stable mode-locking stage. We use a roundtrip circulating-pulse method to simulate the behaviors of beating and stable mode-locking stages [58]. The modeling includes the Kerr effect, the group velocity dispersion of fiber, the saturable absorption of modelocker, and the saturated gain with a finite bandwidth. When the pulses encounter cavity components, we take into account their effects by multiplying the optical field by the transfer matrix of a particular component. The simulation for the roundtrip circulating-pulse method starts with an arbitrary light field with the noise background [e.g., Fig. 7(g)]. After one roundtrip circulation in the laser cavity, the obtained results are used as the input of the next round of calculation until the light field becomes self-consistent. The simulation will approach to a stable solution, which corresponds to a stable laser under certain operation condition. When the optical pulses propagate through the fiber, the extended NSE is used to simulate the dynamics and evolution of the pulses, i.e., [58]

$$\frac{\partial A}{\partial z} + i\frac{\beta_2}{2}\frac{\partial^2 A}{\partial t^2} = \frac{g}{2}A + i\gamma |A|^2 A + \frac{g}{2\Omega_g^2}\frac{\partial^2 A}{\partial t^2}. \quad (C1)$$

Here $A$, $\beta_2$, and $\gamma$ represent the electric filed envelop of the pulse, the fiber dispersion, and the cubic refractive nonlinearity of the fiber, respectively. The variables $t$ and $z$ are the time and the propagation distance, respectively. When the pulses propagate along the SMF, the first and last terms on the right-hand side of Eq. (C1) are ignored. $\Omega_g$ denotes the bandwidth of the gain spectrum. $g$ describes the gain function for the EDF and is given by [58][70]

$$g = g_0 \cdot \exp(-E_p / E_s), \quad (C2)$$

where $g_0$, $E_p$, and $E_s$ are the small-signal gain coefficient related to the doping concentration, the pulse energy, and the gain saturation energy that relies on pump power, respectively. To match the experimental conditions, we use the following parameters: $g_0$=6 dB/m, $\Omega_g$=25 nm, $E_s$=83 pJ, $\gamma$=1.8 W$^{-1}$km$^{-1}$ for EDF, and $\gamma$=1 W$^{-1}$km$^{-1}$ for SMF. Eq. (C1) is solved with a predictor–corrector split-step Fourier method [71].

a movie showing the experimental real-time observation for the formation and evolution of a soliton, including Q-ML, beating dynamics, transient bound state, and stable mode locking.